\documentclass[sn-apa]{sn-jnl}
\usepackage{graphicx}%
\usepackage{xspace}%
\usepackage{multirow}%
\usepackage{amsmath,amssymb,amsfonts}%
\usepackage{amsthm}%
\usepackage{mathrsfs}%
\usepackage[title]{appendix}%
\usepackage{xcolor}%
\usepackage{textcomp}%
\usepackage{manyfoot}%
\usepackage{booktabs}%
\usepackage{algorithm}%
\usepackage{algorithmicx}%
\usepackage{algpseudocode}%
\usepackage{listings}%
\usepackage{fullpage}
\usepackage{setspace}
\usepackage[export]{adjustbox}

\usepackage{attachfile}

\usepackage[normalem]{ulem} % Pacote para comando de riscar texto: \sout{texto}

\newcommand{\ie}{\emph{i.e.}\xspace}
\newcommand{\eg}{\emph{e.g.}\xspace}

\begin{document}

\title[Article Title]{Validating Urban Scaling Laws through Mobile Phone Data: A Continental-Scale Analysis of Brazil's Largest Cities}

%%=============================================================%%
%% Prefix	-> \pfx{Dr}
%% GivenName	-> \fnm{Joergen W.}
%% Particle	-> \spfx{van der} -> surname prefix
%% FamilyName	-> \sur{Ploeg}
%% Suffix	-> \sfx{IV}
%% NatureName	-> \tanm{Poet Laureate} -> Title after name
%% Degrees	-> \dgr{MSc, PhD}
%% \author*[1,2]{\pfx{Dr} \fnm{Joergen W.} \spfx{van der} \sur{Ploeg} \sfx{IV} \tanm{Poet Laureate} 
%%                 \dgr{MSc, PhD}}\email{iauthor@gmail.com}
%%=============================================================%%

\author*[1]{\fnm{Ricardo} \sur{de S. Alencar}}\email{ricardo@lamce.coppe.ufrj.br}

\author[2]{\fnm{Fabiano} \sur{L. Ribeiro}}\email{fribeiro@ufla.br}
\equalcont{These authors contributed equally to this work.}

\author[3,4]{\fnm{Horacio} \sur{Samaniego}}\email{horacio@ecoinformatica.cl}
\equalcont{These authors contributed equally to this work.}

\author[5]{\fnm{Ronaldo} \sur{Menezes}}\email{r.menezes@exeter.ac.uk}
\equalcont{These authors contributed equally to this work.}

\author[1]{\fnm{Alexandre} \sur{G. Evsukoff}}\email{alexandre.evsukoff@coc.ufrj.br}
\equalcont{These authors contributed equally to this work.}

\affil[1]{\orgdiv{COPPE}, \orgname{Federal University of Rio de Janeiro}, \city{Rio de Janeiro}, 
% \postcode{21941-594}, 
\state{RJ}, 
\country{Brazil}}

\affil[2]{\orgdiv{Departamento de Física}, \orgname{Universidade Federal de Lavras}, \city{Minas Gerais}, 
% \postcode{Caixa Postal 3037}, 
\state{MG}, \country{Brazil}}

\affil[3]{\orgdiv{Laboratorio de Ecoinformática, Instituto de Conservación, Biodiversidad y Territorio}, \orgname{Universidad Austral de Chile}, 
% \orgaddress{\street{Campus Isla Teja s/n}, 
\city{Valdivia}, 
% \postcode{5110290}, 
\country{Chile}}

\affil[4]{\orgdiv{Instituto de Sistemas Complejos de Valparaíso}, \city{Valparaíso}, 
% \postcode{7800003}, 
\country{Chile}}

\affil[5]{\orgdiv{Department of Computer Science}, \orgname{University of Exeter}, 
% \city{Exeter}, \postcode{EX4 4QF}, 
\country{United Kingdom}}

%%==================================%%
%% sample for unstructured abstract %%
%%==================================%%

\abstract{Urban scaling theories posit that larger cities exhibit disproportionately higher levels of socioeconomic activity and human interactions. Yet, evidence from developing contexts (especially those marked by stark socioeconomic disparities) remains limited. To address this gap, we analyse a month-long dataset of 3.1~billion voice-call records from Brazil’s 100 most populous cities, providing a continental-scale test of urban scaling laws. We measure interactions using two complementary proxies: the number of phone-based contacts (voice-call degrees) and the number of trips inferred from consecutive calls in distinct locations. Our findings reveal clear superlinear relationships in both metrics, indicating that larger urban centres exhibit intensified remote communication and physical mobility. We further observe that gross domestic product (GDP) also scales superlinearly with population, consistent with broader claims that economic output grows faster than city size. Conversely, the number of antennas required per user scales sublinearly, suggesting economies of scale in telecommunications infrastructure. Although the dataset covers a single provider, its widespread coverage in major cities supports the robustness of the results. We nonetheless discuss potential biases, including city-specific marketing campaigns and predominantly prepaid users, as well as the open question of whether higher interaction drives wealth or vice versa. Overall, this study enriches our understanding of urban scaling, emphasising how communication and mobility jointly shape the socioeconomic landscapes of rapidly growing cities.
}
\keywords{Urban scaling, CDR data, Human mobility, Social interactions}

\maketitle

\onehalfspacing
\section{Introduction}\label{sec:Intro}

Over the years, researchers from diverse areas and disciplines have uncovered the relationship between city population size and various urban metrics \citep{Pumain2004,Kuhnert2006a,bettencourt_growth_2007, ortman_settlement_2015,  loboUrbanScienceIntegrated2020}. While the origin of some of these relationships is scrutinised in urban science \citep{arcaute_constructing_2015,keuschnigg_urban_2019,florida2003cities}, a common empirical characteristic across urban systems is the existence of statistical regularities known as \textit{urban scaling laws} \citep{pumain_evolutionary_2006,bettencourt_growth_2007, ribeiroMathematicalModelsExplain2023}. These regularities remain under discussion \citep{narollPrincipleAllometryBiology1973, ribeiroMathematicalModelsExplain2023, loboUrbanScienceIntegrated2020}, yet they have gained further prominence through integrative theoretical developments spanning biology, physics, and economics \citep{narollPrincipleAllometryBiology1973, Barenblatt1996,West2017,Marquet2005,keuschniggScalingTrajectoriesCities2019}.

Urban scaling laws offer a powerful lens through which to examine how populations interact and shape socioeconomic activities, knowledge production, and infrastructure needs, thereby influencing the social fabric of cities \citep{arbesman_superlinear_2009, bettencourt_invention_2007, nomaler_scaling_2014}. Although such laws have consistently revealed superlinear growth for various urban metrics with city size increases, a significant gap remains in our understanding of the underlying network of human interactions \citep{schlapfer_scaling_2014, samaniegoTopologyCommunicatingCities2020}. Numerous empirical studies have shown a systematic acceleration of social and economic life in larger cities \citep{milgram1970experience, bornstein1976pace,fujita2001spatial, centers2012hiv,bettencourt_why_2008}, yet questions persist about how this acceleration emerges and whether it holds in the context of considerable socioeconomic inequality and heterogeneous urban development.

This discussion resonates with ongoing debates about how scaling provides specific descriptors of system-wide states and dynamics, which can be useful for evaluating long-term persistence, resilience, and tipping points \citep{garmestaniDiscontinuitiesUrbanSystems2008,Levin1999,barnoskyApproachingStateShift2012}. In complex systems such as cities, interactions among components give rise to flows of information, goods, and energy that underlie emergent scaling patterns \citep{ribeiroMathematicalModelsExplain2023}. Yet the conventional mean-field perspective---in which a city is viewed as a homogeneously interacting population---may overlook the reality of loosely connected subsystems and local heterogeneities. Integrating urban scaling with traditional approaches in economics and geography, which focus on the spatial and behavioural intricacies of firms and individuals, will be crucial for elucidating key aspects of underlying urban processes \citep{Glaeser2011, fujitaEvolutionHierarchicalUrban1999, Bettencourt+Samaniego2014, arvidssonUrbanScalingLaws2023}. Such a synthesis would also offer new avenues for sustainably managing complex urban adaptive systems \citep{Levin+Clark2010}.

Recent theoretical work suggests that super-linear scaling patterns may emerge directly from networks of human interactions \citep{arbesman_superlinear_2009, bettencourt_origins_2013, pan_urban_2013}, often tied to scale-invariant increases in per capita social connectivity \citep{bettencourt_origins_2013}. Analyses of mobile phone data, for instance, indicate that while human activity can be highly exploratory, individuals often limit their daily interactions to only a few locations \citep{alessandrettiEvidenceConservedQuantity2018}, reminiscent of the cognitive limit proposed by Dunbar \citep{ruiterDunbarNumberGroup2011}. In spite of these advancements, most studies of human interaction networks in cities rely on indirect assumptions, due to the difficulty of obtaining granular data on the full tapestry of interactions across diverse population groups \citep{Blondel2015}. Furthermore, questions remain over whether widely observed scaling patterns are robust in developing contexts, where inequality, varying infrastructure availability, and uneven technological adoption may alter the mechanisms that link population size to social connectivity.

Still, the accepted notion of increasing returns to scale in bigger cities is often linked to higher densities of interpersonal interaction \citep{ribeiroMathematicalModelsExplain2023,schlapfer_scaling_2014, samaniegoTopologyCommunicatingCities2020}, but limitations in data availability, especially in large developing countries, have impeded conclusive tests of this hypothesis. While certain regions or smaller nations have provided promising results \citep{schlapfer_scaling_2014,samaniegoTopologyCommunicatingCities2020}, few attempts have validated these scaling relationships in contexts characterised by widespread inequality and substantial regional disparities in infrastructure, economic development, and market penetration.

Here, we address this gap by leveraging a Call Detail Records (CDR) dataset of unprecedented scope: 3.1~billion mobile phone records originating from the 100 largest Brazilian cities. Examining such a vast and diverse dataset allows us to test how well urban scaling laws hold in a setting of considerable socioeconomic inequality, and to investigate whether any observed patterns might extend to other rapidly growing regions of the world. We develop two distinct methodological approaches for empirically investigating the relationship between city size and intensity of human interaction. First, we quantify the average number of intra-company mobile phone contacts per person, by city size, expanding on prior empirical investigations in Portugal, the United Kingdom, and Chile \citep{schlapfer_scaling_2014, samaniegoTopologyCommunicatingCities2020}. Second, we supplement this perspective by measuring the number of individual trips across each city size category, hypothesising that actual physical travel can capture an additional, often intangible dimension of urban experience and engagement.

Brazil’s cultural, economic, and geographic diversity provides an ideal laboratory to test the universality of urban scaling. For example, data from the Brazilian Institute of Geography and Statistics (IBGE) in 2020 indicates that São Paulo alone contributes nearly 32\% of the national GDP (approximately US\$500~billion), whereas Manaus, the seventh-largest city, contributes about US\$13~billion. Such disparities not only reflect the high levels of inequality but also raise questions about whether typical scaling patterns can remain valid when confronted with stark variations in infrastructure, technological adoption, and income distribution.

In this article, we examine scaling properties of social interactions---measured both by phone contacts and trips---within these 100 largest Brazilian municipalities. We further address key methodological challenges: correcting market share biases (due to differential penetration of a single operator across cities of varying size), establishing robust frameworks to estimate true interaction levels from a sampled user base, and validating individuals’ presumed residencies via temporal call patterns. Additionally, we identify significant market dynamics in mobile phone adoption, clarifying that adoption rates in larger cities exceed those in smaller localities. Our analyses also reveal new insights into how infrastructure (represented here by antenna deployment) exhibits sublinear scaling with the actual user base, hinting at economies of scale within communication networks.

In summary, the main contributions of this study are as follows. First, we provide a continental-scale validation of urban scaling laws by analysing 3.1~billion mobile phone records across 100 large Brazilian cities, a data scope rarely attained in previous work. Second, we employ a dual approach to measuring human interactions (through both phone contacts and physical trips) demonstrating that super-linear growth emerges under both metrics. Third, we offer methodological refinements for correcting operator market share biases and estimating interaction levels from partially sampled data, while also establishing rigorous residence-validation protocols using temporal call patterns. Fourth, we uncover sublinear scaling of infrastructure when measured against active users rather than against total populations, indicating economies of scale in telecommunications. Fifth, we present evidence of higher mobile usage in bigger cities, underscoring how connectivity itself scales with urban size. Lastly, we show that these patterns persist even in the face of profound socioeconomic inequalities, suggesting that universal mechanisms underlie urban growth processes across a range of social and economic contexts. This perspective extends our understanding of how scaling laws, infrastructure, and human connectivity intertwine in developing nations, highlighting both the power and the adaptability of scaling principles in shaping urban life.

\section{Socio-economics and urban scaling}

Recently, a general framework has emerged connecting socioeconomic metrics $Y$ (\eg, GDP) to interactions between urban dwellers in a city of population size $N$. This framework proposes that:
$Y = g N^2 n_c$
where $n_c$ represents the probability of encounters between urban dwellers, and $g$ denotes the socioeconomic output generated from a single encounter \citep{ribeiroMathematicalModelsExplain2023}. The key distinction among different models describing socioeconomic scaling lies in their estimation of $n_c$. \citet{bettencourt_origins_2013} estimate $n_c$ as the ratio between individually accessible area and total urban infrastructure area. Alternative approaches, including gravity models, conceptualize $n_c$ as the density of contacts:
$n_c = \langle k_i \rangle / N,$
where $\langle k_i \rangle$ represents the average number of contacts between individuals in the city.

Our work adopts this density-based approach. Assuming $g$ is scale-invariant, the socioeconomic output can be expressed as:
$Y \propto N \langle k_i \rangle$.
This formulation reveals that the super-linear scaling of socioeconomic urban metrics emerges directly from the super-linear scaling of contact numbers. Consequently, cities facilitating greater social interactions among urban dwellers tend to generate higher socioeconomic output $Y$. We propose a novel method for computing contact density based on CDR-derived trip data. While motorized trips incur higher costs than telephone calls, necessitating stronger motivation, they serve as robust indicators of significant socioeconomic interactions. This approach is justified by three main factors:

\begin{itemize}
    \item {\bf Cost-benefit considerations:} Physical travel requires substantial investment in time, money, and energy, suggesting these interactions carry higher economic value \citep{aguilera2008business}.
    \item {\bf Interaction quality:} Historically, face-to-face meetings, enabled by physical travel, often facilitate more substantial economic activities than remote communications \citep{storper2004buzz,bathelt2004clusters}.
    \item {\bf Economic activity correlation:} Trip patterns frequently correspond to business activities, professional networking, and consumer behavior, providing direct insights into wealth generation mechanisms \citep{duranton2012urban}.
\end{itemize}

This methodology offers a more nuanced understanding of urban socioeconomic dynamics by capturing high-value interaction densities crucial for economic growth, complementing traditional analytical approaches.

\section{Data and methods}\label{sec:Materials}

\subsection*{Call Detail Records (CDR) data and Brazilian Municipalities} 

Call Detail Records (CDRs) have become a widely used data source for studying human mobility and social interactions, thanks to their high spatial and temporal granularity and widespread coverage \citep{blondel2015survey, ahas2010daily, gonzalez2008understanding}. A CDR typically contains metadata on the origin and destination of a phone call, as well as time stamps and the geographical position of the antenna routing the call. Such data have been leveraged to infer patterns of movement, explore social connectivity, and analyze urban dynamics in numerous contexts.

This study is based on a month-long CDR dataset comprising 3.1~billion records from across the entire Brazilian territory in 2013, covering the period from 13~March to 13~April. The data, provided by a single Brazilian telephone operator, were deliberately selected to represent a typical, non-eventful period, thus minimizing exceptional or seasonal phenomena that might skew estimates of routine social interactions. Voice calls were still widely used at that time, in contrast to the increasing prevalence of voice-over-IP communications in subsequent years. 

%\ronaldo{RM: I think to be stronger we need to have the paper focus on one months but have more months in the SM}
%\textcolor{red}{F2R: But we have available just one month, right?}

The dataset is restricted to calls originating from the operator’s network and does not include additional information such as SMS messages or Internet traffic. To safeguard privacy, the operator fully anonymized the data before sharing it with researchers. A more detailed description of the raw CDR fields, along with exemplar maps of antenna locations and their corresponding Voronoi cells in selected cities, is provided in the Supplementary Material.

Each call is associated with a specific antenna, from which a Voronoi tessellation is constructed to approximate its coverage area (see Figure~\ref{fig:Partitioning by Subdistrict} in the Supplementary Material). Owing to varying antenna densities, these Voronoi polygons can differ significantly in size, leading to decreased positional precision in sparsely covered regions. Consequently, while these partitions offer a practical basis for spatial analysis, researchers should be mindful of potential bias in areas with few antennas.
 
Brazil comprises 5,570 municipalities, ranging from vast megacities with over 10~million residents—supported by numerous antennas—to small towns of only a few thousand inhabitants served by minimal telecom infrastructure. This heterogeneity in urban form, climate, and cultural context makes Brazil an ideal setting for investigating mobility and social interactions at a continental scale. However, analyzing all 5,570 municipalities can dilute the reliability of scaling estimates, particularly for smaller localities lacking robust data. We, therefore, confine our study to the 100 most populous municipalities, whose collective diversity spans multiple climates, cultures, and economic conditions. A full list of these 100 cities and their geographic locations can be found in the Supplementary Material, alongside additional notes on their selection criteria.

Population size for each of these 100 municipalities was obtained from the 2010 census conducted by the Brazilian Institute of Geography and Statistics (IBGE), the nation’s official provider of demographic and socioeconomic statistics \citep{ibge2010censo}. Given the role of population as a core variable in urban scaling analysis, we supplemented the CDR dataset with IBGE-based figures for consistency and comparability. Further details on these census data and our integration methodology are included in the Supplementary Material.

By limiting our analysis to Brazil’s 100 largest cities, we achieve a balance between examining a sufficiently large and diverse sample of urban centers - capturing a multitude of social, cultural, and climatic realities - while still retaining robust statistical power in each locale. This continental-scale investigation allows us to examine whether conventional insights into urban scaling and mobility hold in one of the world’s most heterogeneous and unequal societies. The combination of detailed CDR data and comprehensive census information provides a rich, representative snapshot of everyday mobility and communication, laying the groundwork for subsequent analyses of how city size, infrastructure, and social connectivity interrelate.

\subsection*{Inference of People's Residences}

Understanding individual mobility patterns requires a baseline from which to evaluate daily travel and social interactions, making the identification of users’ residences critical. By establishing where individuals return to or spend their nights, researchers can anchor movement data to a specific ``home'' location. This not only helps differentiate residents from transient visitors but also clarifies commuting behaviors and other regular travel patterns that underlie broader socioeconomic processes.  

We estimate the presumed residence of each user as the antenna location with the highest frequency of calls placed between 7~p.m. and 6~a.m. on weekdays and at any time on weekends. To enhance the accuracy of this estimation, at least 50\% of all calls made during these periods must originate from the identified antenna. Users for whom no clear residential location could be established under these criteria were excluded from subsequent analyses. This approach aligns with prior research efforts that successfully employ mobile phone data to infer home locations~\citep{ahas2010daily,isaacman2011identifying,vanhoofAssessingQualityHome2018}.

\subsection*{Definition of Contacts and Trips}

In order to capture both remote and face-to-face dimensions of social interaction, we adopt two complementary measures derived from the CDR dataset: the number of voice-call \emph{contacts} and the number of \emph{trips}. 
% Although each proxy offers only a partial representation of urban activity, their combined use broadens our understanding of how individuals engage with one another and with their surrounding environment in cities of different sizes.

We define the number of contacts by constructing a complex network whose nodes represent individual users, while the presence of an edge signifies a voice call between two users. Specifically, user \(j\) is considered a \textit{contact} of user \(i\) if user \(i\) has placed one or more calls to \(j\) during the study period. For each node (i.e., user), the total number of distinct contacts can be interpreted as the individual’s effective communication reach via voice calls. This approach follows established network-based analyses of mobile phone datasets, where edges denote social ties inferred from call activity \citep{blondel2015survey, onnela2007structure}.

While voice contacts capture remote communication, they may not fully reflect the face-to-face and economic activities typically requiring physical presence. To account for these, we define a \emph{trip} as the movement between two locations exceeding a minimum distance threshold within a given time window. We set a distance threshold of 2~km occurring within at least 30~minutes and up to 4~hours, thereby reducing the risk of overcounting minor displacements or signal fluctuations in densely populated areas with numerous antennas. As a result, this criterion focuses mainly on medium- to long-range trips commonly associated with work-related journeys, professional engagements, and significant consumer activities, while inevitably under-representing short pedestrian trips.

To operationalise this definition of trips, we develop origin–destination (OD) matrices, following a well-known procedure in the literature \citep{alexander_origindestination_2015,toole_path_2015,lenormand2020entropy,barboza2021identifying}. In particular, we build on the work of \cite{calabrese_understanding_2013}, which infers user movement by examining two consecutive phone calls made from distinct locations (\ie, different antennas). By identifying the origin and destination of each such inferred journey, we form an OD matrix whose entries represent aggregate travel flows between spatial units. More details on the trip detection algorithm, including handling of potential boundary cases and validation steps, are provided in Supplementary material~\ref{app:trip-detection-algo}.

Together, these two metrics —number of contacts and number of trips— serve as proxies for the multifaceted ways in which urban residents interact. Contacts capture communication-based ties, while trips provide insights into physical mobility and face-to-face engagements. Examining both measures in parallel thus allows us to distinguish patterns that may be driven by remote connectivity from those underpinned by in-person encounters.

\section{Results}\label{sec:results}

Results are summarized in Table~\ref{tab:table_empirical_finds}. All urban variables analyzed, including GDP, the number of antennas ($N_A$), and the number of users ($U$), are presented alongside their respective scaling functions and empirical exponents ($\beta$). These exponents quantify how each variable changes with city size, as represented here by the census population or the number of users. Slopes are estimated via ordinary least squares (OLS) regression, and the table also includes measures like standard errors and relevant notes about each variable.
%\textcolor{blue}{The figures in Fig.~\ref{fig:var_v_Pop} and the supplementary material~\ref{append_scaling} display the plots that underpin the results presented in this table.}

\begin{figure}[!h]%
\centering
\includegraphics[width=\textwidth]{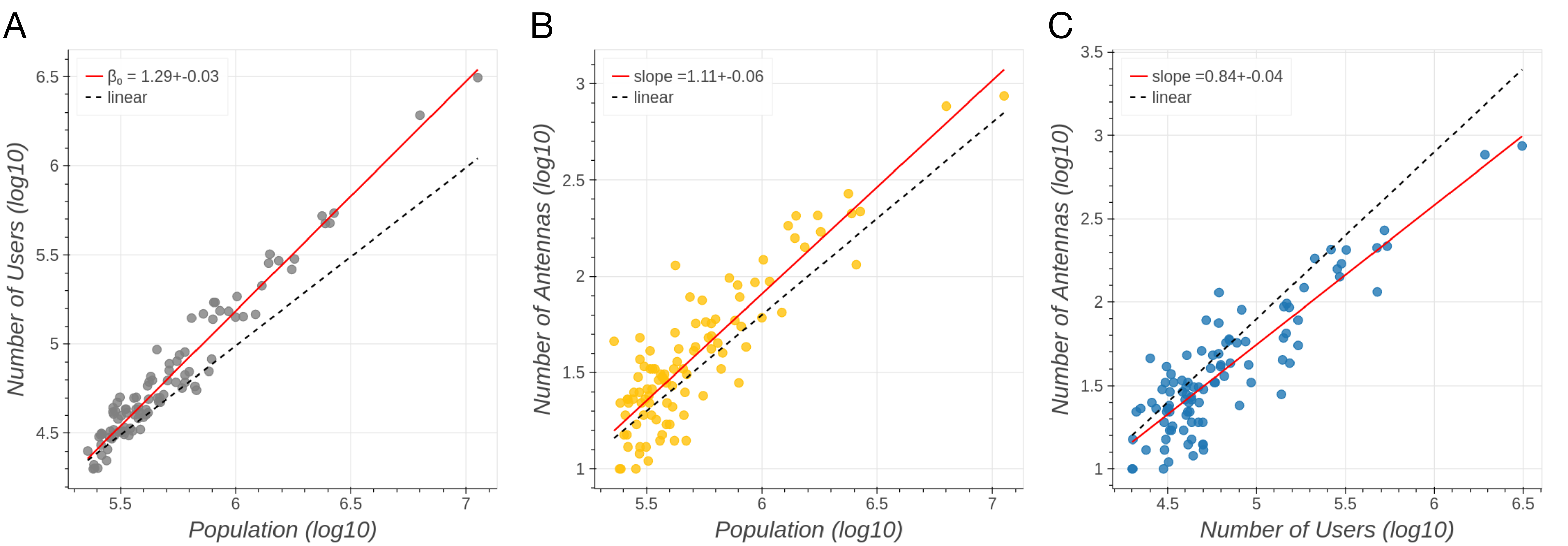}\hfill
\caption{\textbf{Relations between Users ($U$), Population ($P$), and number of antennas ($N_A$).} (A) $U$ exhibits superlinear behavior in relation to $N$. This non-linearity between $U$ and $N$ results in distinct scaling regimes for the number of antennas ($N_A$); (B) superlinear when the number of antennas is considered relative to population size, and (C) sublinear when considered relative to the number of users.
Dashed lines indicate the one-to-one linear relationship, where the slope is equal to 1.0.
}
\label{fig:user_ant_popu}
\end{figure}

Table~\ref{tab:table_empirical_finds} shows a superlinear relationship of GDP with population ($\beta_G = 1.12$). This aligns with previous observations showing that larger cities yield disproportionately higher economic output \citep{lobo_urban_2013}. Additionally, the number of mobile phone users ($U$) grows more steeply ($\beta_0 = 1.29$) than the overall population, indicating that phone adoption intensifies as cities expand. 
As an infrastructural variable, the number of antennas ($N_A$) intriguingly exhibits a superlinear relationship when expressed relative to the population, %($\beta_A = 1.11$), 
but a sublinear relationship when expressed relative to the number of users (Fig.~\ref{fig:user_ant_popu}). % ($\beta_A / \beta_0 = 0.86$). 
%More details about this issue will be discussed in the next subsections. 

The latter part of the table focuses on social interactions (both \emph{contacts} and \emph{trips}) as inferred from call detail records (CDR). Notably, the cumulative degree of the voice-call network shows superlinear scaling with population, implying that larger cities are associated to more contacts per person. A similar trend emerges for trips: the total number of trips in the city, $\tau$, also scales superlinearly with population ($\beta_{\tau} \approx 1.25$), reflecting a heightened level of physical mobility in bigger urban centres. Each of these findings will be discussed in further detail in the following sections, where we explore how they contribute to our broader understanding of urban scaling and social interaction patterns.

%~\ref{tab:table_empirical_finds}.
%\textcolor{red}{F2F:  colocar estas informacoes no appendix, ou entao tiramos a part em azul. }

% \textcolor{red}{Fazer um breve resumo sobre o que vamos apresentar como resultados ....}

\begin{table*}
\centering
\caption{%
Estimated scaling exponents for social interactions and trips, derived from Call Detail Records (CDR) covering Brazil’s 100 most populous cities. 
Social interactions are measured by the number of contacts within the voice-call network, and trips are inferred from users’ call sequences. 
Slopes (\(\beta\)) are obtained via ordinary least squares (OLS) regression. 
\(N\) denotes population size (from demographic data), \(U\) is the number of Users (nodes in the constructed network), and \(T\) corresponds to the number of trips estimated through CDR.
(see Section \ref{sec:Materials})
}
\label{tab:table_empirical_finds}

\resizebox{\textwidth}{!}{%
\begin{tabular}{ll llcc}
    \toprule
    \textbf{Variable} & \textbf{Explanation} 
    & \textbf{Scaling Relation} & \textbf{Value of \(\beta\)} & \textbf{Std.~Error} & \textbf{Observation} \\ 
    \midrule

    $GDP$ 
    & Gross Domestic Product 
    & $GDP \sim N^{\beta_G}$ 
    & $\beta_G = 1.12$ 
    & 0.08 
    & --- \\[6pt]

    $U$ 
    & Number of Users 
    & $U \sim N^{\beta_0}$ 
    & $\beta_0 = 1.29$ 
    & 0.03 
    & --- \\[6pt]

    $U/N$ 
    & Ratio of Users to Population
    & $U/N \sim N^{\beta_0 - 1}$ 
    & $\beta_0 -1 = 0.35$ 
    & 0.03 
    & --- \\[6pt]

    $N_A$ 
    & Number of Antennas 
    & $N_A \sim N^{\beta_A}$ 
    & $\beta_A = 1.11$ 
    & 0.06 
    & \\
    && $N_A \sim U^{\beta_A/\beta_0}$
    & $\beta_A/\beta_0 = 0.86$ 
    & 0.05 
    & Empirical estimation:  \\
    && 
    &
    & 
    & $N_A \sim U^{0.84\pm 0.04}$ (see Fig.\ref{fig:user_ant_popu}c)   \\[6pt]

    $\langle k^s_i \rangle$
    & Avg. Degree in the Sample 
    & $\langle k^s_i \rangle \sim U^{\beta_U}$ 
    & $\beta_U = 0.15$ 
    & 0.01 
    & --- \\[6pt]
    
    $K^s$ 
    & Cumulative Degree in the Sample 
    & $K^s\sim U \langle k^s_i \rangle$ \\
    && $K^s\sim U^{1 + \beta_U}$ 
    & $1 + \beta_U = 1.15$ 
    & 0.01 
    & --- \\[6pt]

    $K$ 
    & Cumulative Degree in the City 
    & $K \sim N^{\beta_c}$ &$ (1+ \beta_0 \beta_U) \to  \beta_c$ (Eq. \ref{Eq_K_teo}) && $\beta_c$ is not \\
    &&& $\beta_c = 1.19$ 
    & 0.02 
    & accessible empirically \\[6pt]

    $\langle T_i^s \rangle$
    & Avg. \# of Trips in the Sample 
    & $\langle T_i^s \rangle \sim U^{\beta_T^s}$ 
    & $\beta_T^s = 0.19$ 
    & 0.04 
    & --- \\[6pt]
    
    $\tau^s$ 
    & Total \# of Trips in the Sample 
    & $\tau^s \sim U \,\langle T_i^s \rangle$ \\
    && $\tau^s \sim U^{1 + \beta_T^s}$ 
    & $1 + \beta_T^s = 1.19$ 
    & 0.04 
    & --- \\[6pt]

    $\tau$ 
    & Total No. of Trips in the City 
    & $\tau \sim N^{\beta_\tau}$ & $ (1+ \beta_0 \beta_T^s) \to \beta_{\tau} $  (Eq. \ref{eq_tau})&& $\beta_{\tau}$ is not \\
    &&& $\beta_\tau = 1.25$ 
    & 0.05 
    & accessible empirically\\
    \bottomrule
\end{tabular}
}
\end{table*}

\subsection*{Number of Phone ID, Population Size, and Adoption Ratios}

Our analysis reveals a superlinear relationship between \(U\) (the number of mobile phones) and \(N\) (the total population), with an exponent \(\beta_0 = 1.29 \pm 0.03\) (Fig.~\ref{fig:user_ant_popu}). This unexpected finding challenges the assumption of a near-linear relationship between population size and mobile phone usage at a large scale. Instead, it indicates that mobile phone adoption increases disproportionately faster than population growth in larger urban areas. 
Specifically, larger cities not only have more residents but also exhibit a higher number of phone IDs per capita. By 2013, most mobile devices supported multiple phone IDs, a trend particularly evident in Brazil, where the prevalence of pre-paid plans often led users to maintain multiple phone IDs across different operators.
One plausible explanation for this phenomenon is that denser, more economically dynamic urban environments foster social and business conditions conducive to the adoption of multiple phone IDs and peer-driven usage patterns. Additionally, these findings suggest the presence of feedback loops in large cities, where the perceived necessity for mobile connectivity accelerates adoption rates. Such feedback mechanisms may influence broader urban dynamics, including the demand for communication infrastructure and the intensity of social interactions, as expanding user bases amplify the value of real-time digital connectivity.

This superlinear behavior may initially appear contradictory to established urban scaling laws, which typically demonstrate that infrastructure grows sublinearly with population. While conventional scaling discussions suggest economies of scale in infrastructure deployment (with exponents ($\beta < 1$)), our findings reveal the opposite pattern for mobile phone antennas ($\beta_0 = 1.11 > 1 \pm 0.06$). This apparent discrepancy highlights an important nuance in interpreting urban scaling: the nominal population count may not always be the most relevant variable for the phenomenon being measured.

In our case, the number of phone IDs ($U$) represents an ``effective population'' for telecommunications, capturing not only the number of individuals but also the intensity and multiplicity of their connectivity needs. When we reframe our analysis with ($U$) as the effective population variable, we observe that the actual telecommunications infrastructure (network antennas) indeed scales sublinearly with this effective user base, consistent with the efficiency predictions of urban scaling theory \citep{bettencourt_growth_2007}. This reconceptualization demonstrates that infrastructure continues to benefit from economies of scale, even as the ``effective user population'' grows superlinearly with the nominal population.
These results underscore the importance of carefully defining population variables in scaling analyzes. For many urban phenomena, particularly those related to technology adoption and usage patterns, the raw count of residents may be less informative than measures of effective participation. This distinction becomes increasingly important in the digital age, where a single individual may maintain multiple digital identities and connection points across various platforms and services.

%
% The initial finding that could have a ripple effect across the other metrics analyzed here is the superlinear relationship between the number of phone company users ($U$), hereafter referred to simply as users,  and population size ($N$), characterized by  $U \sim N^{1.29}$, which means scaling exponent  $\beta_0 = 1.29 \pm 0.03$, as shown in table~\ref{tab:table_empirical_finds} and Figure \ref{fig:var_v_Pop}.
%
% For instance, the scaling of the ratio of users to the total population size exhibits a scaling behavior where $U/N \sim N^{0.29}$, indicating that our dataset captures a larger market share in among larger cities.

%%% INCLUDE IN DISCUSSION!!!
% \textcolor{blue}{
% Surprisingly, this finding contrasts with Schläpfer et al. \cite{schlapfer_scaling_2014}, who report that market share in the countries they analyzed (Portugal and the UK) does not change significantly with increasing city size.}
% The superlinearity in the number of users observed in Brazilian cities could have two potential explanations. It may represent an emergent scaling property or simply reflect the company’s increased activity and aggressiveness in larger urban areas. Additionally, we cannot entirely rule out the possibility of the dataset underestimating the number of users in smaller cities.
%%%

\begin{figure}[!h]
    \centering
% %    \includegraphics[width=.5\textwidth]{Imagens/scalings_panel_vert.pdf}
%  a) 
%  \includegraphics[width=.35\textwidth]{Imagens/users-x-pop.png}
% b) 
% \includegraphics[width=.35\textwidth]{Imagens/market-share-x-N.png}
% \\ c) 
\includegraphics[width=.35\textwidth]{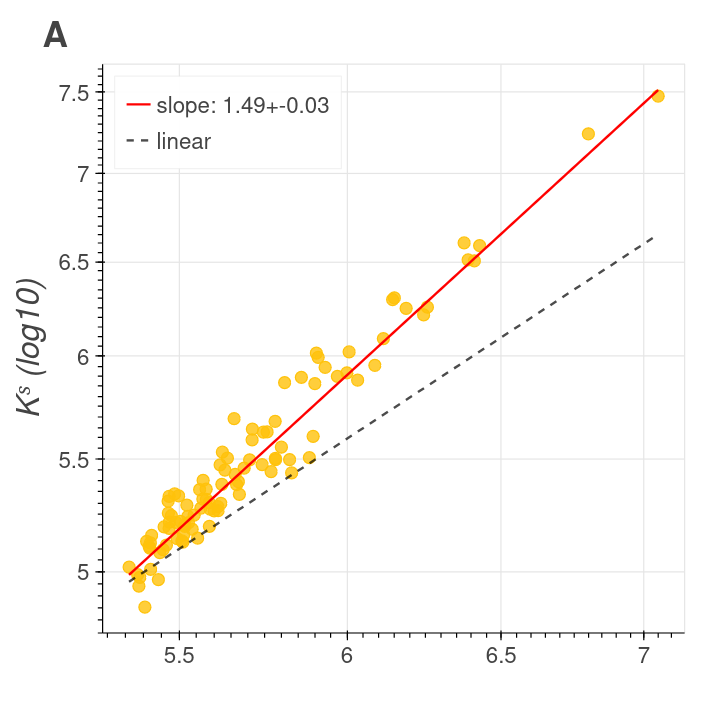}
\includegraphics[width=.35\textwidth]{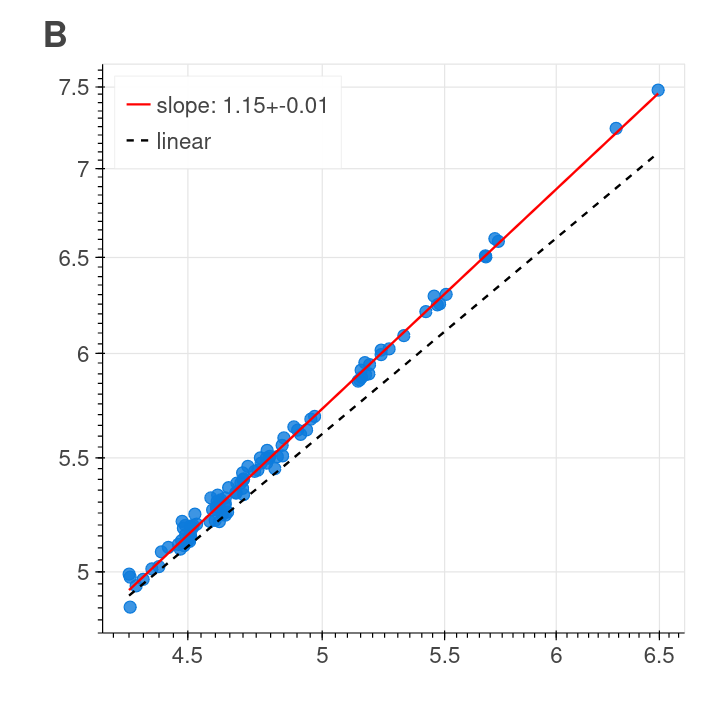}
\\ 
\includegraphics[width=.35\textwidth]{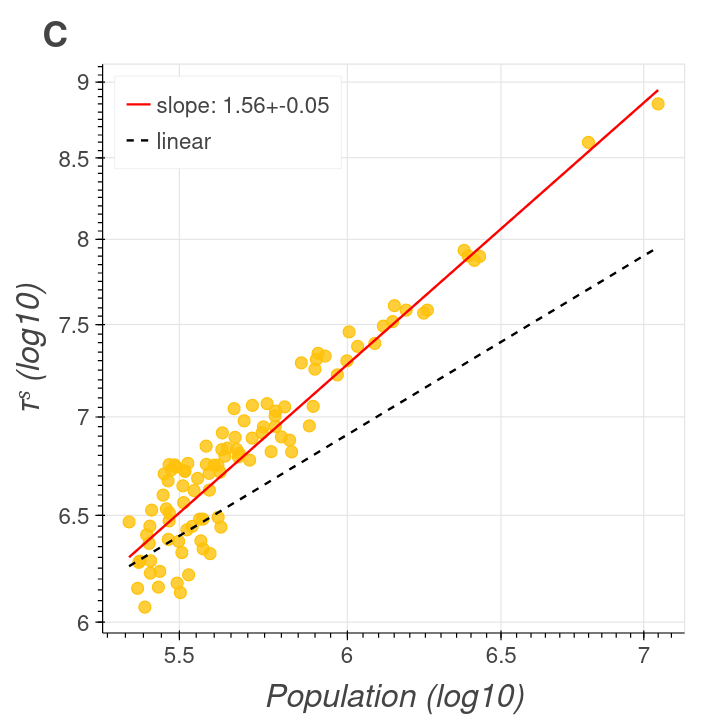}
\includegraphics[width=.35\textwidth]{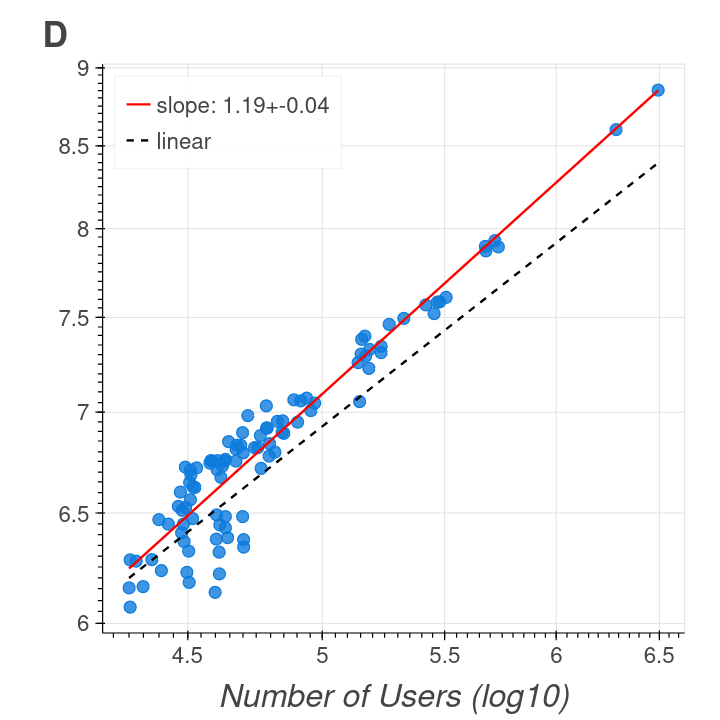}
  \caption{
\textbf{Scaling of degree and trips.} (Top): Cumulative degree in the dataset ($K^s$) as a function of (A) city population size ($N$) and (B) the number of users ($U$).
(Bottom): Number of trips restricted to the dataset ($\tau^s$) as a function of (C) population size and (D) the number of users.
Note that the superscript 's' represents the ``sample'' estimation. Dashed lines indicate the one-to-one linear relationship, where the slope is equal to 1.0.}
\label{fig:var_v_Pop}
\end{figure}

\subsection*{Number of Antennas}

While the number of antenna among cities may be viewed as an infrastructure variable that usually shown sublinear scaling with population size \citep{bettencourt_invention_2007,meirelles_evolution_2018}, our work shows that the number of antennas (\(N_A\)) is superlinear (\(\beta_A = 1.11 \pm 0.06\)), implying that larger populations are served by a disproportionately larger number of antennas.
In contrast, when using the number of mobile phone users ($U$) as a proxy for city scale, the number of antennas becomes sublinear ($N_A \sim U^{0.86}$) matching observations from the literature for infrastructural urban features.

Several explanations may account for the observed superlinear scaling of the number of antennas ($N_A$) with respect to population size ($N$). One plausible explanation is that larger cities, characterized by more diverse and complex activity patterns, may require operators to deploy additional antenna sites to ensure adequate service quality and coverage. Alternatively, this trend could reflect the tendency of operators to prioritize or aggressively expand infrastructure in larger cities, where the potential for higher returns on investment and user demand is greater.
Regarding the sublinear relationship between the number of antennas ($N_A$) and the number of users ($U$), one possible explanation is that denser clusters of users enable more efficient utilization of shared antenna capacity, reducing the need for proportional infrastructure expansion and improving cost-effectiveness. Additionally, operators may strategically prioritize antenna deployment based on actual device usage patterns rather than relying solely on city population size.
This observation supports the hypothesis that companies adjust their infrastructure investments in response to user demand dynamics rather than population metrics alone. Furthermore, the sublinear scaling of $N_A$ with respect to $U$ may reflect economies of scale in network deployment —an effect that is not evident when considering population size ($N$) as the primary variable. This suggests that user density and demand play a critical role in shaping infrastructure investment strategies, potentially leading to more efficient resource allocation in urban environments.

Collectively, these findings highlight how the choice of city-size metric can significantly influence our understanding of infrastructure scaling dynamics. When measured relative to total population ($N$), more populous cities exhibit an accelerated rate of antenna deployment. However, when assessed relative to the number of mobile users ($U$), larger user populations can be served with proportionally fewer new antennas, reflecting efficiencies in network utilization.

This result underscores the importance of carefully selecting metrics in scaling analyses. Since the number of users scales non-linearly with population size, applying scaling methods based solely on population metrics may yield misleading predictions of scaling exponents. Consequently, incorporating user-based metrics is essential to accurately capture the underlying dynamics of infrastructure deployment and avoid erroneous conclusions.

% Another finding relates to the number of antennas, denoted as $N_A$. The number of antennas is an urban infrastructure variable and, as per the categorization proposed in \citep{bettencourt_invention_2007}, it should scale sublinearly with population size, indicating an economy of scale.
% However, our analysis reveals that \(N_A \sim N^{1.11}\) (i.e., $\beta_A = 1.11 \pm 0.06$), indicating a superlinear scaling with city population size.
%
% Interestingly, when we analyze the scaling not in terms of population size but in terms of the number of users, we find 
% $N_A \sim U^{0.86}$ (i.e. $\beta_A/\beta_0 = 0.86 \pm 0.05$). That is,  a sublinear relationship between the number of antennas and the number of users, suggesting that larger cities require fewer antennas per user.
%
%% 
%%% I'ld move the following to the discussion
% This finding leads to the hypothesis that companies adjust the deployment of new antennas based on the number of users rather than solely on city population size, possibly aligning with demand dynamics.
% \textcolor{blue}{
% In addition, this finding illustrates the need for caution in our scaling analysis. Given that the number of users scales non-linearly with population size, naively applying scaling methods based on population size could lead to incorrect predictions of the true value of the scaling exponent.}

\subsection*{Social Interaction Estimations}
\label{sec:num_contacts}

\subsubsection*{Number of Mobile Phone Contacts}

Finally, we evaluated how social interactions scale with population size by examining city-level communication networks derived from voice-call interactions. Our goal is to estimate the level of interaction in cities of various sizes, specifically through the average number of mobile phone contacts per person (\(\langle k_i \rangle\)) and the total number of contacts across the entire urban population (\(K\)).

To formalize these metrics, let \(k_i\) denote the number of contacts (i.e., degree) of an individual \(i\). Then, the average degree in a city of population \(N\) is

\begin{equation}
\langle k_i \rangle = \frac{1}{N} \sum_{i=1}^N k_i,
\end{equation}
and multiplying by the total population yields the citywide cumulative degree:
\begin{equation}\label{Eq_K}
    K = N \,\langle k_i \rangle.
\end{equation}
Computing these quantities directly would require knowing \(k_i\) for every single inhabitant, which is unfeasible. However, Call Detail Records (CDRs) allow us to estimate  \(k_i\) by tracking the number of distinct mobile phone contacts associated with each user in our sample. Specifically, we define
\begin{equation}
\langle k_i^s \rangle = \frac{1}{U} \sum_{i=1}^U k_i^s,
\end{equation}
where \(k_i^s\) is the number of voice-call contacts of the sampled user \(i\) with other sampled users residing in the same city, and \(U\) is the total number of sampled users. Consequently,
\begin{equation}\label{Eq_Ks}
    K^s = U \,\langle k_i^s \rangle
\end{equation}
gives the cumulative degree restricted to the subset of users for whom CDR data are available.

Note that \(\langle k_i^s \rangle\) and \(K^s\) can be directly computed from the data, but \(\langle k_i\rangle\) and \(K\) for the full population of size \(N\) must be inferred. We assume 

\begin{equation}\label{eq_ki_kis}
 \langle k_i^s \rangle \approx \langle k_i \rangle \, ,   
\end{equation}
following arguments also made by \cite{schlapfer_scaling_2014}, under the rationale that an individual’s propensity to maintain contacts does not depend strongly on the share of users captured in the sample.

We observed a noteworthy discrepancy in the exponents obtained when scaling \(K^s\) with \(U\) versus scaling \(K^s\) with \(N\). Namely, we find \(\beta \approx 1.15\) when expressing \(K^s\) in terms of \(U\), versus \(\beta \approx 1.49\) with respect to \(N\). We attribute this difference to non-linearities in the relationship between \(U\) and \(N\).
To reconcile these metrics, we use the assumption in Eq.~\ref{eq_ki_kis} along with Eqs.~\ref{Eq_K} and~\ref{Eq_Ks} to express the cumulative degree of the entire city in terms of the cumulative degree of the sample:

\begin{equation}
    K \sim \left(\frac{N}{U}\right)\,K^s,
\end{equation}
which, in turn, leads us to
\begin{equation}\label{Eq_K_teo}
    K \sim N^{1 + \beta_0 \beta_U},
\end{equation}
given the exponents defined in Table~\ref{tab:table_empirical_finds}. Plugging in the empirically obtained \(\beta_0 = 1.29\) and \(\beta_U = 0.15\) yields \(K \sim N^{1.19}\), signalling a clear superlinear regime. Consequently, the average number of contacts per person follows \(\langle k_i \rangle \sim N^{0.19}\), which means it is larger in larger cities.  

To illustrate, in São Paulo (\(\sim 10^7\) inhabitants), the average user maintains roughly 10 mobile phone contacts, whereas in Belém (\(\sim 1.3\times10^6\) inhabitants), this figure drops to about 7 contacts. %Appendix~\ref{appendix_degree_distr} provides further insight into the distribution of \(k_i\) values in Brazilian cities, revealing a near log-normal profile. 
This pattern aligns with earlier studies in  Portugal and Chile \citep[\ie][]{schlapfer_scaling_2014, samaniegoTopologyCommunicatingCities2020},
which also find superlinear growth in the number of social contacts. However, our results expand the literature by applying these methods to a continental-scale dataset spanning highly diverse socioeconomic and geographic conditions, reinforcing the universality of superlinear interaction growth across larger urban populations.

\subsubsection*{Number of Trips}\label{sec:num_trips}

To complement our analysis of phone-based contacts, we also examine the number of trips within a city as an alternative proxy for urban interaction. In many cases, physical travel more directly captures face-to-face meetings and location-based activities, offering insights into the economic and social dimensions of everyday life that may not be fully reflected by remote communication alone. Similarly to the preceding section on contacts, we define an average number of trips per person, \(\langle T_i \rangle\), as follows:

\begin{equation}
\langle T_i \rangle = \frac{1}{N} \sum_{i=1}^N T_i,
\end{equation}
where \(T_i\) is the number of trips undertaken by the \(i\)-th city resident over the specified observation period. From this, we obtain the total number of trips \(\tau\) in the city by multiplying \(\langle T_i \rangle\) by the population \(N\):

\begin{equation}\label{eq_tau}
    \tau = N \langle T_i \rangle,
\end{equation}
analogous to the cumulative degree.

In practice, \(\langle T_i \rangle\) and \(\tau\) are not directly observable from raw CDR data. We therefore rely on a sampling strategy, analogous to the approach used for contact estimation. Specifically, let \(T_i^s\) denote the number of trips made by the user \(i\) in our sample of size \(U\). We can then directly compute:

\begin{equation}
\langle T_i^s \rangle = \frac{1}{U} \sum_{i=1}^U T_i^s,
\end{equation}
which in turn yields the total trips in the sample:
\begin{equation}\label{eq_taus}
    \tau^s = U \langle T_i^s \rangle.
\end{equation}

As before, \(\tau\) for the entire city is unavailable, so we estimate it by assuming \(\langle T_i \rangle \approx \langle T_i^s \rangle\). This assumption follows the rationale that each individual’s trip behavior should not depend on the fraction of the population captured in the sample. Combining this assumption with Eqs.~\eqref{eq_tau} and \eqref{eq_taus}, we get

\begin{equation}
    \tau \sim \left(\frac{N}{U}\right) \tau^s,
\end{equation}
and by incorporating the scaling relations described in Table~\ref{tab:table_empirical_finds}, we obtain
\begin{equation}\label{eq_tau_scaling}
    \tau \sim N^{1 + \beta_0 \beta_T^s}.
\end{equation}

Substituting the empirically determined exponents \(\beta_0 = 1.29 \pm 0.03\) and \(\beta_T^s = 0.19 \pm 0.04\), we arrive at \(\tau \sim N^{1.25}\) (\(\beta_{\tau} = 1.25 \pm 0.05\)). This result indicates a clear superlinear relationship between the total number of trips and population size, mirroring the pattern observed for phone-based contacts. In other words, larger cities do not simply have more trips; they have a disproportionately higher volume of collective travel, suggesting that increased physical mobility may be reinforcing the same agglomeration dynamics observed in communication networks. Further discussion of the implications and contrasts between trips and contacts can be found in Section~\ref{sec:Discussion}, where we consider how these two facets of urban interaction jointly shape socioeconomic outcomes.

\section{Discussion}
\label{sec:Discussion}

In this work, we analyze the scaling laws of human interaction by comparing urban indicators to social variables derived from a large sample Call Detail Records (CDRs) derived from mobile phone interactions.
Our analyses consistently suggest that as city size increases, so does the \emph{per capita} number of contacts and trips. In essence, people living in bigger urban centers tend to interact more frequently (both via phone calls and physical travel) than those in smaller ones. According to scaling theory \citep{bettencourt_origins_2013, ribeiroMathematicalModelsExplain2023}, this arises because the \emph{probability of encounters} grows with population density, creating agglomeration effects \citep[\eg][]{Duranton2004a,Duranton2014} and boosting thereby opportunities for social, economic, and cultural exchanges. 

We employ two proxies to quantify human interaction: the average number of mobile phone contacts ($\langle k_i \rangle$) and the number of trips ($\langle T_i \rangle$). While mobile phone contacts reflect remote communication, trips capture activities dependent on physical movement. Our analysis reveals that both metrics exhibit similar superlinear scaling trends, supporting the hypothesis that larger cities foster more intense interaction patterns. Although each proxy has limitations—mobile phone contacts may overlook face-to-face interactions, while trips may not account for brief or virtual exchanges—their combined use offers a more comprehensive and multidimensional understanding of urban connectivity.

Both measures -contacts and trips- scale superlinearly and may contribute to enhanced socio-economic performance in bigger cities. In line with other urban scaling studies, our results indicate that either more extensive contact networks or greater mobility (or both) could feed into wealth creation processes. However, it remains unclear whether heightened social connectivity is the cause or a consequence of urban wealth. Future research should investigate the feedback mechanisms through which vibrant local economies attract more residents, thereby increasing interaction rates and further stimulating economic growth.

While our findings demonstrate consistent patterns, we recognize several potential biases in the data. First, call patterns exhibit significant temporal variation, with pronounced peaks during working hours. Second, our dataset predominantly consists of prepaid SIM cards (roughly 85\% at the time of data collection), which are often associated with lower-income demographics and may reflect lower call frequencies. Furthermore, short pedestrian trips are likely underrepresented, as they may not meet the distance and time thresholds required for trip detection. To address these limitations, we aggregate features (e.g., \ average contacts or average trips) at the city level. However, residual biases may still affect the precise values of our scaling exponents.

We also highlight that the causal relationship between high interaction rates and economic wealth remains unresolved: does a high volume of social interactions drive economic productivity, or does existing wealth facilitate a greater frequency of interactions? Addressing this question will require more controlled experiments and refinements to existing theoretical frameworks.

Ultimately, a comprehensive understanding of urban scaling processes —encompassing both social and economic dimensions— will necessitate a deeper examination of factors contributing to GDP beyond basic demographic data \citep[see][]{lobo_urban_2013, leiScalingUrbanEconomic2021}. This could include longitudinal call detail record (CDR) datasets spanning multiple years, as well as mobility data such as that presented in this study. Such approaches will be critical for disentangling the complex interplay between social dynamics, infrastructure, and economic outcomes in urban systems.

Our findings demonstrate that both social contacts and mobility patterns scale superlinearly with population size, offering a potential explanation for the enhanced socioeconomic performance observed in larger cities compared to smaller ones. However, further research is needed to disentangle the causal relationships underlying these patterns and to assess their generalizability across diverse urban systems. Such efforts will be critical for advancing our understanding of the mechanisms driving urban scaling and their implications for socioeconomic outcomes

A final limitation stems from the fact that our data are sourced from a single mobile phone provider. While this operator has a sufficiently large customer base to ensure strong coverage in the 100 largest Brazilian cities, factors such as city-specific factors such as marketing campaigns, promotional pricing, or infrastructure investments could introduce bias. To mitigate this, we focused on a single, non-eventful month, reducing the likelihood that short-term promotions disproportionately influenced specific locations. Consequently, while our results suggest that larger cities indeed display higher mobile adoption and more active phone usage, verifying these findings with data from multiple providers, or over a longer timeframe, would help confirm their robustness and generality.

\section{Conclusion}

Quantitative analysis of how human interactions and the resulting agglomeration effects reverberates in wealth creation represents one of the most significant advancements enabled by big data analytics. 
Prior to this, particularly in the works of \cite{marshall1890growth} and \cite{Jacobs1969,jacobs1961life}, such understanding was limited to insights and, at best, qualitative hypotheses.
However, with the widespread availability of vast datasets —such as those derived from the continuous use of mobile phones— the science of cities approaches the status of an exact science. 
Urban theories can now be empirically tested and falsified, much like the methodologies that have underpinned physical sciences for over three centuries.

Our findings reveal an empirical scaling relationship between the average and total number of mobile phone contacts within a city and its population size. Specifically, we observe that larger cities tend to support a higher volume of social interactions.
This study provides a substantial contribution to the understanding of how human interactions are measured and, to the best of our knowledge, represents the first large-scale effort to empirically validate urban scaling laws.

%As principais contribuições deste trabalho são:
%A organização de uma base de dados das 100 maiores cidades brasileiras contendo os dados de CDR, dados socioeconômicos, mapas e arquivos georeferenciados;
%A estimativa do número médio de viagens mensais para cada uma das 100 cidades;
%A estimativa do número médio de contatos na mesma cidade para cada uma das cidades;
%A utilização do número médio de viagens e contatos como proxy do número de interações num modelo de lei de escala;
%Os resultados demostram que os modelos propostos são razoavelmente aderentes aos dados observados e estão alinhados com outros resultados apresentados na literature.

\section*{Declarations}

For the purpose of open access, the author has applied a Creative Commons Attribution (CC BY) licence to any Author Accepted Manuscript version arising from this submission.

\section*{Acknowledgements}
HS was supported by Fondecyt (grant number \#1211490) form the chilean national R\&D agency (ANID) and COPPE.
F.\ L.\ Ribeiro thanks CNPq (grant numbers 403139/2021-0 and 424686/2021-0) and Fapemig (grant number APQ-00829-21 and APQ-06541-24) for financial support. Alexandre G. Evsukoff was supported by CNPq (grant number 303503/2021-1).

% Some journals require declarations to be submitted in a standardised format. Please check the Instructions for Authors of the journal to which you are submitting to see if you need to complete this section. If yes, your manuscript must contain the following sections under the heading `Declarations':

% \begin{itemize}
% \item Funding 1211490
% \item Conflict of interest/Competing interests (check journal-specific guidelines for which heading to use)
% \item Ethics approval 
% \item Consent to participate
% \item Consent for publication
% \item Availability of data and materials
% \item Code availability 
% \item Authors' contributions
% \end{itemize}

% \noindent
% If any of the sections are not relevant to your manuscript, please include the heading and write `Not applicable' for that section. 

\backmatter

%%===================================================%%
%% For presentation purpose, we have included        %%
%% \bigskip command. please ignore this.             %%
%%===================================================%%
\bigskip

\begin{appendices}

\section{CDR Dataset} \label{app:cdr-definition}

The study was conducted using data provided for research purposes by a telecommunications operator. The mobile telephony database comprises 30 days of call records from the year 2013, covering the period between March 21, 2013, and April 19, 2013, totaling 3.1 billion records spanning the entire Brazilian territory.
Only data from calls made by users of the telecommunications operator were made available for the study. As a result, the dataset does not contain additional traffic information, such as incoming calls from other operators, text messages (SMS), or internet traffic.
Call location information was inferred from the location data (Latitude and Longitude) of the antennas that processed the calls. The database contains only records of outgoing calls, meaning it is not possible to infer the location of users who received calls.

As shown in Table~\ref{table_descriptionCDR}, a Call Detail Record (CDR) carries information related to the date, time, and duration of the calls, as well as details about the area code (Direct Distance Dialing – DDD) and the phone numbers of both the caller and the recipient. To preserve user privacy, phone number information is encrypted. Additionally, the CDR contains details about the type of traffic used in the call (such as international roaming) and the names of both the originating and destination operators.

\begin{table}[!h]
\centering
\caption{ \label{table_descriptionCDR}
Data description of one particular call contained in the telephone call records (CDR).}
\begin{tabular}{p{.15\textwidth}p{.6\textwidth}}
\toprule
\textbf{Field} & \textbf{Description} \\ 
\midrule
day & The day of the record. \\
time & The time of the record. \\ 
duration & The duration of the call. \\ 
ddd\_orig & The area code of the station originating the call. \\ 
num\_orig & The encrypted ID identifier of the originating station of the call. \\  
ddd\_dest & The station DDD of the destination. \\ 
num\_dest & The encrypted ID of the destination station for the call. \\ 
cell\_id lat & Latitude coordinate of the antenna that processed the call at the origin \\ 
cell\_id long & Longitude coordinate of the antenna that processed the call at the origin \\  
tp\_traffic & Type of the call, local, international, roaming etc. \\  
hold\_orig & Name of the operator that processed the call at the origin. \\  
hold\_dest & Name of the operator that handled the call at the destination. \\  
\bottomrule
\end{tabular}
\end{table}

\section{The 100 most populous Brazilian municipalities }

\begin{figure}[!h]%
\centering
\includegraphics[width=.8\textwidth]{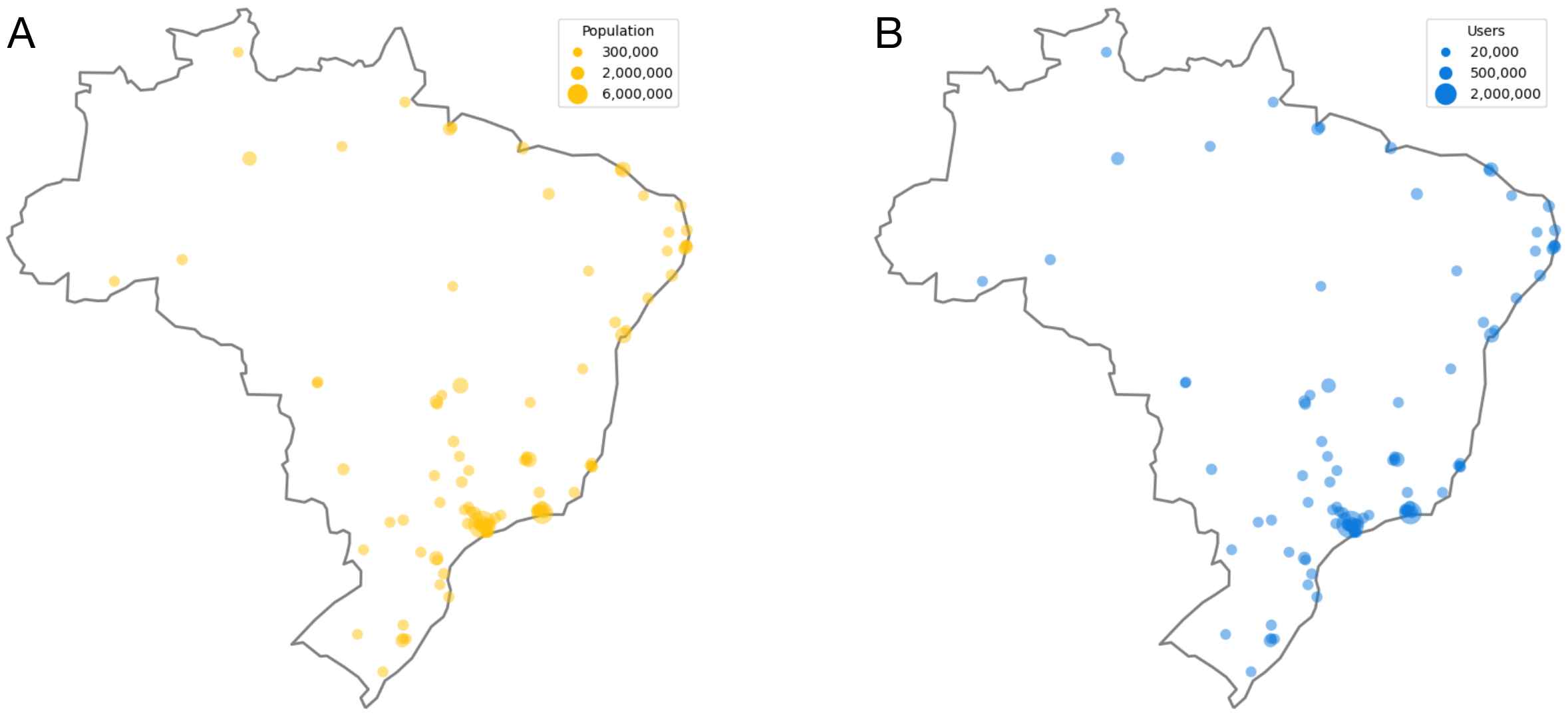}
\caption{\textbf{The 100 most populous municipalities in Brazil analysed in this study.} (A) node size represents the population of these 100 cities whereas on the (B), the node size represents the number of users. While visually they seem to correlate, the number of users ($U$) grows superlinearly with population ($P$) (Figure \ref{fig:user_ant_popu}(Left)).}
\label{fig:map_100cities}
\end{figure}

Brazil is home to 5,570 municipalities, ranging from large megacities with over 10 million people to small towns with only a few thousand residents. This diversity in urban form, climate, and culture makes Brazil an ideal location to study mobility and social interactions. However, analyzing all municipalities could reduce the reliability of scaling estimates, especially for smaller cities with limited data. Therefore, the study focuses on the 100 most populous municipalities, which span a wide range of climates, cultures, and economic conditions. 
Population data for these cities was obtained from the 2010 IBGE census, ensuring consistency with CDR datasets. By focusing on these 100 cities, the study balances diversity with statistical robustness.
The 100 most populous municipalities analyzed in this study are represented on the map shown in Fig.~\ref{fig:map_100cities}.

\section{Scaling Analysis}
\label{append_scaling}

\begin{figure}[!h]%
\centering
\includegraphics[width=\textwidth]{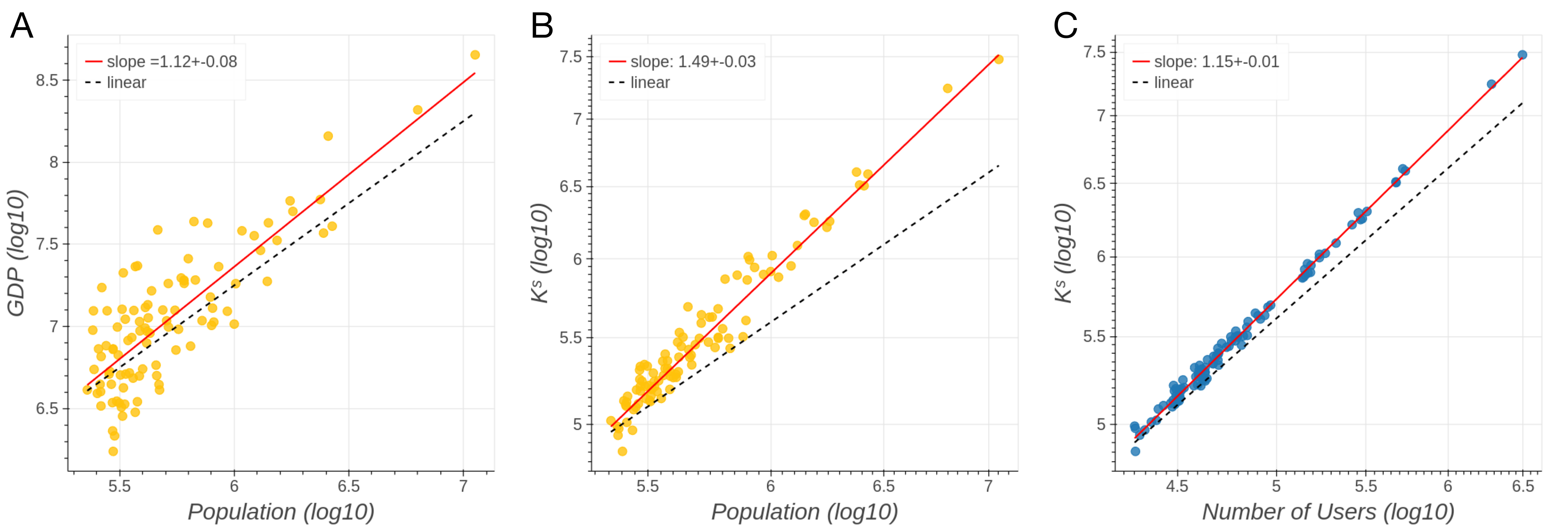}
\caption{
\textbf{Scaling analysis of various metrics and estimation of the scaling exponent.}
Graphs like the ones shown here give rise to the scaling exponents described in Table~\ref{tab:table_empirical_finds}.
Panel (A) presents the superlinear scaling between GDP and population size.
The scaling of cumulative degree ($\langle K^s \rangle$) relative to population size and the number of users is shown in panels (B) and (C), respectively.
Note that the \textit{axis} have been log-transformed.}
\label{fig:scaling_variables}
\end{figure}

The scaling analysis conducted in this paper examines how a given metric, denoted as $Y$, scales with a reference metric, such as population size or the number of users.
Empirical evidence suggests \cite{bettencourt_growth_2007} that a power-law relationship of the form

\begin{equation}\label{eq_app_power} 
Y \sim N^{\beta} 
\end{equation}
holds when city population size ($N$) is used as the scaling parameter. In this study, we assume that a similar power-law relation also applies when the number of users ($U$) is considered as the scaling parameter.
In the equation above, $\beta$ is the scaling exponent that we aim to estimate. To do so, we first take the logarithm of Eq.~\ref{eq_app_power} and then apply ordinary least squares (OLS) regression to estimate the exponent.
The estimated scaling exponents for various urban metrics are presented in Table~\ref{tab:table_empirical_finds}. Additionally, Figures~\ref{fig:user_ant_popu}, \ref{fig:var_v_Pop}, and \ref{fig:scaling_variables} illustrate the scaling behavior of some of these urban variables.

\section{Trip detection algorithm}\label{app:trip-detection-algo}

The methodology for estimating the OD matrix, considered for calculating trips in each municipality, counts trips without distinguishing whether the user departed from home or not.
The algorithm described below presents the steps used to estimate the OD matrices from the CDR data used in this study.  The calculation only considers users with an identified presumed address. Trip detections are then performed, where it is verified whether two consecutive calls from the same ID are made. By ordering ID, DAY, TIME, it is observed whether there was a change in the geographic position (latitude and/or longitude) of the antenna that processed the next call.

%A user's trip is determined by the distance traveled in a given time.
The main idea of trip estimation in this work is to capture motorized trips. A trip is considered a trip if a minimum of 2 km was "traveled" in a minimum of 30 minutes and a maximum of 4 hours. The distance "traveled" between consecutive calls is the linear distance between the telephone towers that processed the calls. In other words, it is an approximate distance, taking into account that the user makes calls within the antenna coverage area and that the trip route is different from a straight line.
%The scale factor e, which refers to a population expansion factor, serves the purpose of uniformly expanding the telephone operator's "market share" in the study area, that is, in the 100 most populous municipalities. In this work, as the purpose of calculating trips is to use them in the analysis of scaling laws, so that the weight of the population does not bias the effect of the scale in relation to the population, a scale factor of 1 was used, that is, the number of trips and the number of users were maintained without applying expansion factors.
%So that the analysis can be carried out based on total quantities for the cities, 
After obtaining the OD matrices for each municipality, the number of trips is obtained from the sum of all trips.

\begin{algorithm}
\caption{OD matrix estimation from CDR data}

\textbf{Input: } CDR logs, the constants $\Delta T$ and $L_{min}$, table of distances between geographic units and $T_{min}T_{max}$ \\
\textbf{Output: } OD Matrix
\begin{algorithmic}[1]
\State \textbf{Begin}
\State $OD$ = NULL
    
\State\indent \textbf{For} each trip detected by two successive records of the same user, within distance, $l_{ij}$, and time interval, $\Delta T$
\State \indent \indent \textbf{If} $l_{ij}>\ L_{min}$ and $T_{min}\le \Delta T\le T_{max}$
\State \indent \indent \indent  ${OD}_{ij}\leftarrow {OD}_{ij}$ + $e_i$
\State \indent \indent \textbf{End If}
\State \indent \textbf{End For}
\State Returns $OD$ Matrix
\State \textbf{End}
\end{algorithmic}
\end{algorithm}

\section{The Voronoi cells}

\begin{figure}[ht]%
\centering
\includegraphics[width=\textwidth]{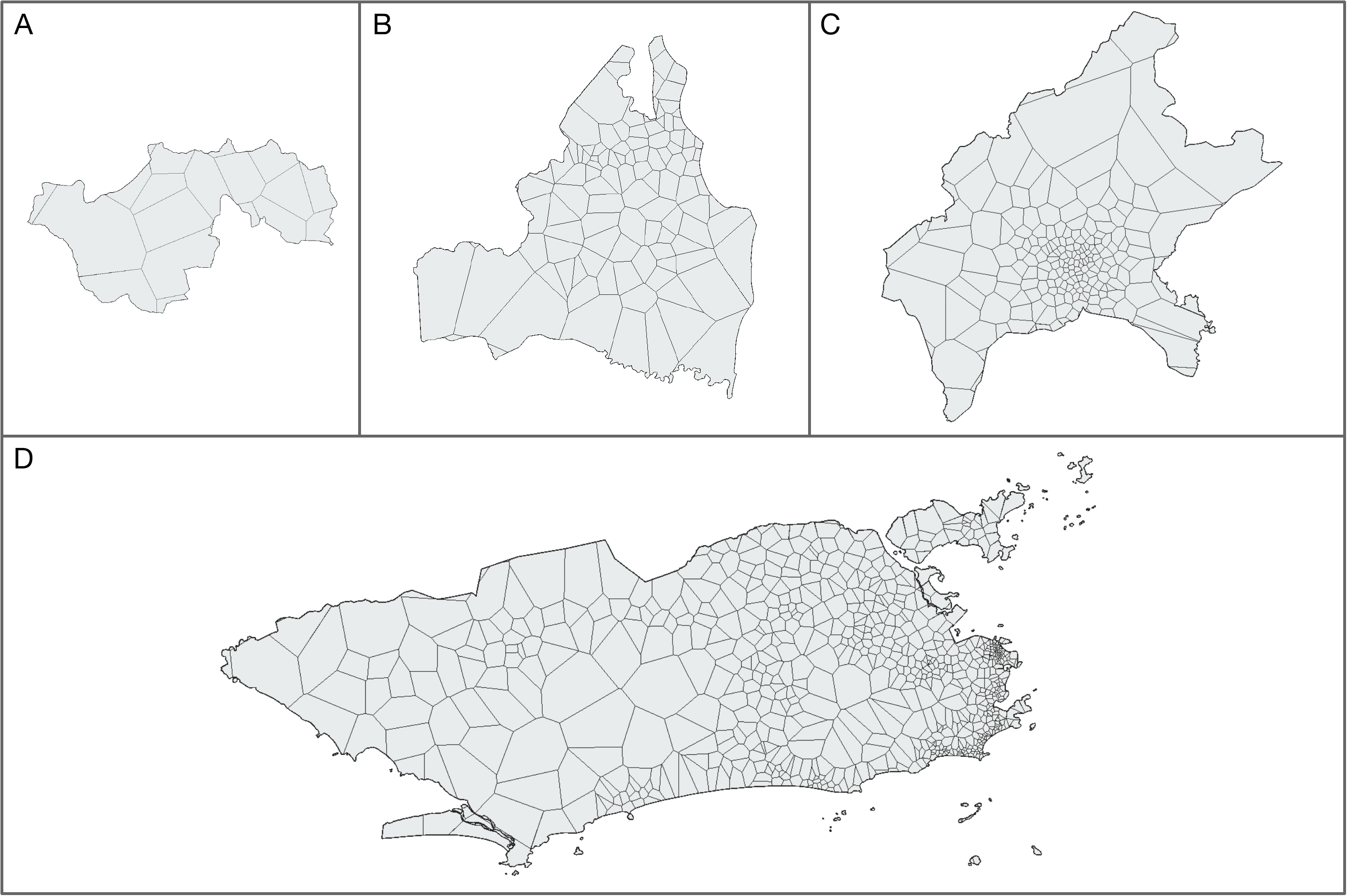}
\caption{\textbf{Voronoi tessellation of cellular network coverage in 4 Brazilian cities.} (A) Sumaré, SP (population $\approx$241K). (B) João Pessoa, PB (population $\approx$723K). (C) Goiânia, GO (population $\approx$1.3M). (D) Rio de Janeiro, RJ (population $\approx$6.3M). Despite antenna placement prioritizing population density and socioeconomic factors, these visualizations reveal significant heterogeneity in spatial coverage across cities of varying sizes.}
\label{fig:Partitioning by Subdistrict}
\end{figure}

Each call in the CDR dataset is associated with a specific antenna, from which a Voronoi tessellation is constructed to approximate its coverage area, as illustrated for 4 cities of different sizes in \figurename~\ref{fig:Partitioning by Subdistrict}. Due to varying antenna densities, these Voronoi polygons can differ significantly in size, leading to decreased positional accuracy in sparsely covered regions. Consequently, while these partitions provide a practical basis for spatial analysis, it is important to be aware of potential bias in areas with fewer antennas.

\section{Aggregated data}

The processed and aggregated data for the 100 largest cities in Brazil, including various urban variables (such as fractal dimension, total street length, number of traffic-related deaths, among others), are contained in the spreadsheet provided in the supplementary material.

\end{appendices}

\bibliography{bibliography}% common bib file
%% if required, the content of .bbl file can be included here once bbl is generated
%%\input sn-article.bbl

\end{document}